\documentclass{article}

\usepackage{amsmath}
\usepackage{amsfonts}
\usepackage{amssymb}

\def\be{\begin{eqnarray}}
\def\ee{\end{eqnarray}}
\def\nn{\nonumber}

\hoffset= - 1.0in         

\title{{\bf Proving AGT relations in the large-c limit
} \vspace{.2cm}}
\author{{\bf A.Mironov}\footnote{ {\small {\it
Lebedev Physics Institute} and {\it ITEP, Moscow, Russia}};
mironov@itep.ru; mironov@lpi.ru} \ and {\bf
A.Morozov}\thanks{{\small {\it ITEP, Moscow, Russia}};
morozov@itep.ru} \date{ }}

\begin{document}

\maketitle

\vspace{-5.0cm}

\begin{center}
\hfill FIAN/TD-24/09\\
\hfill ITEP/TH-44/09\\
\end{center}

\vspace{3.cm}

\begin{abstract}
In the limit of large central charge $c$ the 4-point Virasoro
conformal block becomes a hypergeometric function. It is represented
by a sum of chiral Nekrasov functions, which can also be explicitly
evaluated. In this way the known proof of the AGT relation is
extended from special to generic set of external states, but in
the special limit of $c=\infty$.
\end{abstract}

\section{Introduction}

The AGT relations \cite{AGT}-\cite{AGTlast} express conformal blocks
\cite{BPZ,Zams,CFT} of $2d$ chiral algebras through the Nekrasov functions
\cite{Nek}-\cite{Neklast}. In the case of the Virasoro block with
$4$ primaries, the both sides of the relation depend on $6$ free parameters: five
dimensions, four "external" and one "internal" which we parameterize as
\be
\Delta_i =
\frac{\alpha_i(\epsilon-\alpha_i)}{\epsilon_2\epsilon_2}, \ \ \ \ i=
0,\ldots,4, \label{dims}
\ee
and the central
charge, parameterized as $c = 1 + \frac{6\epsilon^2}{\epsilon_1\epsilon_2}$,
$\epsilon= \epsilon_1+\epsilon_2$. The relation states that
\be
\sum_{|Y|=|Y'|} x^{|Y|} \gamma_{\Delta\Delta_1\Delta_2}(Y)Q_\Delta^{-1}(Y,Y')
\gamma_{\Delta\Delta_3\Delta_4}(Y') =
(1-x)^{-\nu}\sum_{Y,Y'} x^{|Y|+|Y'|}
Z_{\Delta;\Delta_2\Delta_2;\Delta_3\Delta_4}(Y,Y')
\label{agt}
\ee
For notations and other details see \cite{MMMagt}. The sum goes over pairs
of Young diagrams, but in two different ways: it is diagonal in the
number of boxes, $|Y|=|Y'|$ at the l.h.s., while the summation variables are
totally free (unconstrained) at the r.h.s.
These two expansions are related to boson and fermion representations of
more general $\tau$-functions \cite{gentau}, what deserves a more detailed
study and discussion.
In fact, there are plenty of different questions about the AGT relations,
which connect the transcendental and often controversial field of Seiberg-Witten
theory \cite{SW} and integration over singular instanton moduli spaces
with the basic group theory and complex analysis, unified  into a difficult
but well defined subject of $2d$ conformal field theory.

In \cite{Gnc,mmm2,mmm3} the two limiting cases of (\ref{agt}) were considered:
one of large external dimensions, which on the Nekrasov-SW side corresponds to
the case of non-conformal (asymptotically free) SYM models,
and the other one of large internal dimension $\Delta_0$, where the nice
Zamolodchikov asymptotic formula \cite{Zam} allows one to effectively deal
with the old
controversial case \cite{Khrev} of the instanton calculus in
$4d$ conformal
invariant model with $N_f=2N_c$ (one can confirm that instanton corrections exist
and even odd numbers of instantons contribute, moreover, the end-point of RG flow
is described by an elegant modular relation, at least, for $N_c=2$).

This letter is devoted to one more limit, $c\rightarrow\infty$.
In this limit, either $\epsilon_1\rightarrow 0$ or $\epsilon_2\rightarrow 0$.
Then only the {\it chiral} Nekrasov functions, i.e.those with $(Y,Y') =
([1^n],\emptyset)$ or $(\emptyset,[1^n])$ contribute to the r.h.s. of (\ref{agt}),
while the l.h.s. becomes a hypergeometric series.
In other words, the limit reproduces the situation studied in \cite{mmNF}
and \cite{mmU3}, where the AGT relations were {\it proved} (this is the only case
where a complete {\it explicit} proof already exists) for the Fateev-Litvinov conformal blocks
\cite{FLit}. The difference is that there restricting the hypergeometricity and
chirality came from a {\it special} selection of external states, while
here it is enough to take, say, $\epsilon_{1}\rightarrow 0$ without constraining
external states.

\section{Hypergeometric conformal block}

The fact that
\be
B_{\Delta;\Delta_1\Delta_2\Delta_3\Delta_4}(x)
\ \stackrel{c\rightarrow\infty}{\longrightarrow}\
\phantom._2F_1\Big(\Delta+\Delta_1-\Delta_2,
\Delta+\Delta_3-\Delta_4;2\Delta;x\Big) = \nn \\ =
\sum_{n=0}^\infty \frac{x^n}{n!}\prod_{k=0}^{n-1}
\frac{(\Delta+\Delta_1-\Delta_2+k)
(\Delta+\Delta_3-\Delta_4+k)}{2\Delta+k}
\label{Fhser}
\ee
is known since \cite{Zac}.
Still it deserves reminding a simple derivation.

It is instructive to begin with the first terms
of the expansion. It can be taken from any standard text-book on $2d$ conformal
field theory \cite{CFT,BPZ}, in order to have all the notations consistent,
we use below \cite{MMMagt}. The conformal block is defined in a highly
asymmetric way and depends on the order of dimensions.
Explicitly,
\be
B(x) = 1 + x\frac{(\Delta+\Delta_1-\Delta_2)
(\Delta+\Delta_3-\Delta_4)}{2\Delta} +
\nn
\ee
\be\label{B2t}
+x^2
\left[
{(\Delta+\Delta_1 -\Delta_2   )(\Delta+\Delta_1 -\Delta_2   +1)
(\Delta+\Delta_3-\Delta_4)
(\Delta+\Delta_3-\Delta_4+1)\over 4\Delta(2\Delta+1)}+\right.\ee\be\left.
+{\left[(\Delta_2   +\Delta_1 )(2\Delta+1)+\Delta(\Delta-1)
-3(\Delta_2   -\Delta_1 )^2\right]
\left[(\Delta_3+\Delta_4)(2\Delta+1)+\Delta(\Delta-1)
-3(\Delta_3-\Delta_4)^2\right]
\over 2(2\Delta+1)\Big(2\Delta(8\Delta-5) + (2\Delta+1)c\Big)}\right]
+
\nn
\ee
\be
+\ldots
\nn
\ee
The linear in $x$ term does not depend on $c$ at all,
and the $x^2$-term depends  only through a single
entry, $c/2$ in the Shapovalov matrix. For large
$c$, inverse of this matrix has a single non-vanishing
element,
\be
Q^{-1} \ \stackrel{c \rightarrow \infty}{\longrightarrow}\
\left(\begin{array}{cc} 0 & 0 \\ 0 & \frac{1}{4\Delta(2\Delta+1)}
\end{array}\right)
\ee
Thus, in this limit (\ref{B2t}) becomes
\be
B(x)  \ \stackrel{c \rightarrow \infty}{\longrightarrow}\
1 + x\frac{(\Delta+\Delta_1-\Delta_2)
(\Delta+\Delta_3-\Delta_4)}{2\Delta} + \nn \\ +\frac{x^2}{2}
\frac{(\Delta+\Delta_1-\Delta_2)(\Delta+\Delta_1-\Delta_2+1)
(\Delta+\Delta_3-\Delta_4)
(\Delta+\Delta_3-\Delta_4+1)}{2\Delta(2\Delta+1)} + \ldots
\ee
where one can easily recognize the first terms of the
hypergeometric series (\ref{Fhser}).

In the generic $x^n$-term the same thing happens:
the only elements of the Shapovalov matrix that do not
involve $c$-dependent Virasoro commutators, thus being
independent of $c$ and, hence, not growing with $c$,
are $\langle L_{-1}^n\Delta| L_{-1}^n \rangle\ =\
\langle \Delta| L_1^nL_{-1}^n |\Delta\rangle$.
Therefore, in general as $c\rightarrow\infty$,
\be
B(x) = \sum_{n=0}^\infty x^n \frac{
\langle L_{-1}^nV_\Delta|V_{\Delta_1}V_{\Delta_2}\rangle
\langle L_{-1}^nV_\Delta|V_{\Delta_3}V_{\Delta_4}\rangle}
{\langle L_{-1}^n\Delta|L_{-1}^n\Delta\rangle}
\ee
Note that the limit to pure gauge theories \cite{Gnc,mmm2} is similarly
controlled by the $L_{-1}^n$-contributions only.
However, particular contributions and the resulting formulas
are essentially different from those in the large-$c$ limit.

The triple vertices in the numerator are known from \cite{BPZ,Zams,Bel,MMMagt,MMMM}:
\be
\langle L_{-1}^nV_\Delta|V_{\Delta_1}V_{\Delta_2}\rangle =
\prod_{k=0}^{n-1} (\Delta+\Delta_2-\Delta_2+k)
\ee
and the second vertex is just the same, with $(1,2)\rightarrow(3,4)$
(the last property is special for the Virasoro symmetry, it is not true
in general, see \cite{MMMM,mmU3}).
The matrix element in the denominator is found
by the recursive procedure:
\be
L_1L_{-1}^n|\Delta\rangle = L_{-1}L_1L_{-1}^{n-1}|\Delta\rangle
+ 2(\Delta+ n-1)L_{-1}^{n-1}|\Delta\rangle =\!\!
\sum_{j=1}^n 2(\Delta + n-j) L_{-1}^{n-1}|\Delta\rangle=
n(2\Delta + n-1)L_{-1}^{n-1}|\Delta\rangle
\ee
and
\be
L_1^nL_{-1}^n|\Delta\rangle =
n(2\Delta + n-1)L_1^{n-1}L_{-1}^{n-1}|\Delta\rangle
= n!\prod_{k=0}^{n-1} (2\Delta+ k) |\Delta\rangle
\ee
This completes the proof of (\ref{Fhser}).

\section{Chiral Nekrasov functions}

In the limit $\epsilon_1\rightarrow 0$ not only $c\rightarrow\infty$,
but also the dimensions (\ref{dims}) grow infinitely:
$\Delta_i \rightarrow \infty$.
In order to avoid this one needs to especially adjust
$\alpha$-parameters: choose them so that $\alpha_i\sim\epsilon_1$. Then, (\ref{dims})
implies that
\be
\Delta_i={\alpha_i (1 + O(\epsilon_1))\over \epsilon_1}\ ,\ \ \ \ \hbox{i.e.}\ \ \ \
\alpha_i = \epsilon_1\Delta_i + O(\epsilon_1^2)
\label{alim}
\ee
This behavior of $\alpha_i$, in particular, means that one can neglect the
$U(1)$-factor $(1-x)^{-\nu}$ in (\ref{agt}), because
typically $\nu = \frac{2\alpha_1\alpha_3}{\epsilon_1\epsilon_2} = O(\epsilon_1)$.

\subsection{The $x$-linear terms}

We begin with the illustration: the $x$-linear term at the r.h.s.
of (\ref{agt}):
\be
x\Big(Z_{[1][0]}+Z_{[0][1]}\Big) =
-x\left(\frac{\prod_{f=1}^4 (a+\mu_i)}{\epsilon_1\epsilon_2\ 2a(2a+\epsilon)}
+ \frac{\prod_{f=1}^4 (-a+\mu_i)}{\epsilon_1\epsilon_2\ 2a(2a-\epsilon)}\right)
\label{Z1}
\ee
where parameters $a$ and $\mu_i$ are related to $\alpha_i$ linearly \cite{AGT,MMMagt}:
\be
a = \alpha_0 - \frac{\epsilon}{2}, \ \ \
\mu_1=-{\epsilon\over 2}+\alpha_1+\alpha_2, \ \ \mu_2={\epsilon\over
2}+\alpha_1-\alpha_2, \ \ \mu_3=-{\epsilon\over
2}+\alpha_3+\alpha_4, \ \ \
\mu_4={\epsilon\over 2}+\alpha_3-\alpha_4,
\label{agtrel}
\ee

\subsubsection{The simplest case: $\alpha_{1,2,3,4}=0$}

If we now put $\alpha_{1,2,3,4}=0$, then (\ref{Z1}) becomes
\be
-x\frac{a^2-\epsilon^2/4}{2\epsilon_1\epsilon_2} =
x\frac{\alpha_0 (\epsilon-\alpha_0)}{2\epsilon_1\epsilon_2}
=\frac{\Delta_0}{2}
\ee
This is in excellent match with the conformal block side
of (\ref{agt}).

We even did not use (\ref{alim}).
What is important for us here, however: only one of the two $Z$-functions
actually contributes to (\ref{Z1}) in this limit.
Indeed,
\be
Z_{[1][0]} = -\frac{(a^2-\epsilon^2/4)^2}{\epsilon_1\epsilon_2\
2a(2a+\epsilon)} = -\frac{\alpha_0^2(\alpha_0+\epsilon)^2}
{2\epsilon_1\epsilon_2\ 2\alpha_0(2\alpha_0-\epsilon)}
= \frac{\Delta_0}{2} + O(\epsilon_1)
\ee
if $\alpha_0$ satisfies (\ref{alim}), while
\be
Z_{[0][1]} = -\frac{(a^2-\epsilon^2/4)^2}{\epsilon_1\epsilon_2\
2a(2a-\epsilon)} = -\frac{\alpha_0^2(\alpha_0+\epsilon)^2}
{\epsilon_1\epsilon_2 (2\alpha_0-\epsilon)(2\alpha_0-2\epsilon)}
=  O(\epsilon_1)
\ee

\subsubsection{$\alpha_{1,2,3,4}\ne 0$}

This remains true when $\alpha_{1,2,3,4}$ are switched on.
Then, from AGT relations (\ref{agtrel}) it follows that
\be\label{masses1}
\underline{a+\mu_1 = \alpha_0 + \alpha_1-\alpha_2}, \nn \\
a + \mu_2 = \alpha_0 + \alpha_1+\alpha_2 - \epsilon, \nn \\
\underline{a+\mu_3 = \alpha_0 + \alpha_3-\alpha_4}, \nn \\
a + \mu_2 = \alpha_0 + \alpha_3+\alpha_4 - \epsilon, \nn \\
\nn \\
-a+\mu_1 = -\alpha_0 + \alpha_1-\alpha_2+\epsilon, \nn \\
\underline{-a + \mu_2 = -\alpha_0 + \alpha_1+\alpha_2}, \nn \\
-a+\mu_3 = -\alpha_0 + \alpha_3-\alpha_4+\epsilon, \nn \\
\underline{-a + \mu_2 = -\alpha_0 + \alpha_3+\alpha_4},
\ee
Underlined are the quantities that vanish in the limit of $\epsilon_1\rightarrow 0$,
combined with the prescription (\ref{alim}). Now one immediately obtains that
\be
Z_{[0][1]} =-\frac{\epsilon^2\epsilon_1^2(\Delta_0-\Delta_1-\Delta_2)
(\Delta_0-\Delta_3-\Delta_4)+O(\epsilon_1)}
{\epsilon_1\epsilon_2 (2\alpha_0-\epsilon)(2\alpha_0-2\epsilon)}
=O(\epsilon_1)
\ee
and
\be
Z_{[1][0]} =  -\frac{\epsilon^2\epsilon_1^2(\Delta_0+\Delta_1-\Delta_2)
(\Delta_0+\Delta_3-\Delta_4)+O(\epsilon_1)}
{\epsilon_1\epsilon_2 2\alpha_0(2\alpha_0-\epsilon)}
= \frac{(\Delta_0+\Delta_1-\Delta_2)
(\Delta_0+\Delta_3-\Delta_4)}{2\Delta} + O(\epsilon_1)
\ee
which coincides with (\ref{B2t}).
Note that of eight factors (\ref{masses1})
only the first and third ones provide $\Delta$-dependent contributions
to the conformal block in the limit of interest.

\subsection{The $x^2$ terms}

In this case, there are five Nekrasov functions:
\be
{\cal Z}_{[2][0]} = \frac{1}{2!\,\epsilon_1\epsilon_2^2
(\epsilon_1-\epsilon_2)}\cdot
\frac{\prod_{r=1}^4 (a + \mu_r)(a+\mu_r+\epsilon_2)}
{2a(2a+\epsilon_2)(2a+\epsilon)(2a+\epsilon+\epsilon_2)},\\
{\cal Z}_{[0][2]} = \frac{1}{2!\,\epsilon_1\epsilon_2^2
(\epsilon_1-\epsilon_2)}\cdot
\frac{\prod_{r=1}^4 (a - \mu_r)(a-\mu_r-\epsilon_2)}
{2a(2a-\epsilon_2)(2a-\epsilon)(2a-\epsilon-\epsilon_2)},\\
{\cal Z}_{[11][0]} = -\frac{1}{2!\,\epsilon_1^2\epsilon_2
(\epsilon_1-\epsilon_2)}\cdot
\frac{\prod_{r=1}^4 (a + \mu_r)(a+\mu_r+\epsilon_1)}
{2a(2a+\epsilon_1)(2a+\epsilon)(2a+\epsilon+\epsilon_1)},  \\
{\cal Z}_{[0][11]} = -\frac{1}{2!\,\epsilon_1^2\epsilon_2
(\epsilon_1-\epsilon_2)}\cdot
\frac{\prod_{r=1}^4 (a - \mu_r)(a-\mu_r-\epsilon_1)}
{2a(2a-\epsilon_1)(2a-\epsilon)(2a-\epsilon-\epsilon_1)}, \\
{\cal Z}_{[1][1]} = \frac{1}{\epsilon_1^2\epsilon_2^2}\cdot
\frac{\prod_{r=1}^4 (a + \mu_r)(a-\mu_r)}
{(4a^2-\epsilon_1^2)(4a^2-\epsilon_2^2)};
\label{Z2}
\ee
In order to estimate them, one also needs to complement
(\ref{masses1}) by
\be\label{masses2}
a+\mu_1 + \epsilon_2 = \alpha_0 + \alpha_1-\alpha_2 + \epsilon_2, \nn \\
\underline{a + \mu_2 +\epsilon_2 = \alpha_0 + \alpha_1+\alpha_2 - \epsilon_1}, \nn \\
a+\mu_3 +\epsilon_2 = \alpha_0 + \alpha_3-\alpha_4 + \epsilon_2, \nn \\
\underline{a + \mu_2 +\epsilon_2 = \alpha_0 + \alpha_3+\alpha_4 - \epsilon_1},
\ee
where again underlined are the quantities that vanish in the limit of $\epsilon_1\rightarrow 0$,
combined with prescription (\ref{alim}).

Combining these formulas with prescription (\ref{alim}) and the AGT relations
(\ref{agtrel}), one obtains
$$
{\cal Z}_{[2][0]} = O(\epsilon_1)
$$
$$
{\cal Z}_{[0][2]} = O(\epsilon_1)
$$
$$
{\cal Z}_{[11][0]} = -\frac{1}{2!\,\epsilon_1^2\epsilon_2
(\epsilon_1-\epsilon_2)}\ \cdot
$$
\be
\cdot\
\frac{\epsilon^4\epsilon_1^4(\Delta_0+\Delta_1-\Delta_2)
(\Delta_0+\Delta_3-\Delta_4)(\Delta_0+\Delta_1-\Delta_2+1)
(\Delta_0+\Delta_3-\Delta_4+1)+O(\epsilon_1)}
{(2\alpha_0-\epsilon)(2\alpha_0-\epsilon_2)2\alpha_0(2\alpha_0+\epsilon_1)}=\\
=\frac{1}{2!}\cdot
\frac{(\Delta_0+\Delta_1-\Delta_2)
(\Delta_0+\Delta_3-\Delta_4)(\Delta_0+\Delta_1-\Delta_2+1)
(\Delta_0+\Delta_3-\Delta_4+1)}
{2\Delta_0(2\Delta_0+1)}+O(\epsilon_1)
,
\ee
$$
{\cal Z}_{[0][11]} =O(\epsilon_1^2)
$$
$$
{\cal Z}_{[1][1]} =O(\epsilon_1)
$$
which reproduces with (\ref{B2t}).

\subsection{Selection rule in the general case}

In the generic case, factors in the numerator that are small in the limit under
consideration are underlined in the lists (\ref{masses1}) and (\ref{masses2}).
To each of them one can also add any number of $\epsilon_1$, but {\it not}
$\ \epsilon_2$.
Similarly, in the denominator the factors
\be
2a + \epsilon_2 + k\epsilon_1 = 2\alpha_0 + k\epsilon_1
\ee
are the only vanishing in this limit among all $2a + k\epsilon_1+l\epsilon_2$.

\paragraph{Nekrasov functions.}
At the next step we need an explicit  expression for generic $Z_{Y,Y'}$:
\be
{\cal Z}_{Y,Y'}={\eta(Y,Y')\over \xi(Y,Y')}
\ee
where for any {\it ordered} pair of Young diagrams $Y$ and $Y'$
\be\label{etaU}
\eta(Y,Y')=\prod_{(i,j)\in Y}\prod_{\alpha=1}^4
\Big(\phi(a_1,i,j)+\mu_\alpha\Big)
\prod_{(i',j')\in Y'}\prod_{\alpha=1}^4
\Big(\phi(a_2,i',j')+\mu_\alpha\Big)
\ee
\be\label{xidef}
\xi(Y,Y')=\prod_{(i,j)\in Y}E(a_1-a_1,Y,Y',i,j)
\Big(\epsilon-E(a_1-a_1,Y,Y,i,j)\Big)\times\nn\\
\times\prod_{(i,j)\in Y}E(a_1-a_2,Y,Y',i,j)
\Big(\epsilon-E(a_1-a_2,Y,Y',i,j)\Big)\times\nn\\
\times\prod_{(i',j')\in Y'}E(a_2-a_1,Y',Y,i',j')
\Big(\epsilon-E(a_2-a_1,Y',Y,i',j')\Big)\times\nn\\
\times\prod_{(i',j')\in Y'}E(a_2-a_2,Y',Y',i',j')
\Big(\epsilon-E(a_2-a_2,Y',Y',i',j')\Big),
\ee
$a_2=-a_1$ and
\be
\phi(a,i,j)=a+\epsilon_1(i-1)+\epsilon_2(j-1)
\label{phi}\\
E(a,Y,Y',i,j)=a+\epsilon_1\Big(k^T_j(Y)-i+1\Big)-\epsilon_2\Big(k_i(Y')-j\Big)
\ee

\bigskip
\noindent
We now need to find all the elementary factors in this product, which vanish as
$\epsilon_1, \alpha_i \rightarrow 0$.

\paragraph{Numerator.}
With the numerator it is simple:
we should find all the terms of the form $\prod(a+\mu_f+k\epsilon_1)$,
$\prod(a+\mu_f+\epsilon_2+k\epsilon_1)$ and $\prod(-a+\mu_f+k\epsilon_1)$,
i.e. those which have either $i=1$ or $i=2$ or $i'=1$ in (\ref{phi}).
Each such product
contributes $\epsilon_1^2$, thus the numerator goes to zero as
\be
\eta(Y,Y') \sim \epsilon_1^{2[\#(i=1)+\#(i=2)+\#(i'=0)]}
\ee

\paragraph{$a$-independent contribution to the denominator.}
The denominator contains factors of two different types:
independent of $a$ and dependent on $a$.
Contributions of the first type come from the first and the last line in (\ref{xidef}).
The factor $E(0,Y,Y,i,j) = \epsilon_1\Big(k_j^T - i+1\Big) - \epsilon_2\Big(k_i(Y)-j\Big)$
does not contain $\epsilon_2$ and thus vanishes when $\epsilon_1\rightarrow 0$:
if $j=k_i(Y)$, i.e. when the box $(i,j)\in Y$ is at the left edge of a row.
The number of such elements is exactly the same as the number of rows, i.e.
is equal to $k_1^T(Y) = \#(l=0)$.
Another factor in the first line of (\ref{xidef}) is  $\epsilon -E(0,Y,Y,i,j) =
-\epsilon_1\Big(k_j^T - i\Big) + \epsilon_2\Big(k_i(Y)-j+1\Big)$, it would vanish
as $\epsilon_1\rightarrow 0$ if $j=k_i(Y)+1$, i.e. if the box is beyond the diagram.
Thus, this factor never vanishes in the limit of interest.
Taking into account both diagrams $Y$ and $Y'$, the
contribution of the $a$-independent terms to denominator is
$\sim\epsilon_1^{\#(i=1) + \#(i'=1)}$.

\paragraph{$a$-dependent contribution to the denominator.}
Contributions which depend on $a$ come from the second and third lines in (\ref{xidef}).
The factor $E(2a,Y,Y',i,j) = 2a+\epsilon_1\Big(k_j^T(Y)-i+1\Big) -
\epsilon_2\Big(k_i(Y')-j\Big)$
vanishes in the limit iff the coefficient of $\epsilon_2$
is unity, i.e. iff $j = k_i(Y')+1$. This can happen whenever the heights of rows
of the two diagrams are such that $k_i(Y)\geq k_i(Y')+1$.
Similarly,
$\epsilon-E(2a,Y,Y',i,j) = -2a-\epsilon_1\Big(k_j^T(Y)-i\Big) +
\epsilon_2\Big(k_i(Y')-j+1\Big)$ vanishes in the limit when the coefficient
in front of $\epsilon_2$ is $-1$, i.e. if $j=k_i(Y')+2$, what can happen
whenever $k_i(Y) \geq k_i(Y')+2$.
The two other factors from the third line can be analyzed in the same way:
$E(-2a,Y',Y,i',j') = -2a+\epsilon_1\Big(k_j^T(Y')-i'+1\Big) -
\epsilon_2\Big(k_i(Y)-j'\Big)$ vanishes in the limit iff
$k_i(Y)-j'=1$, i.e. $j=k_i(Y)-1$, what is possible whenever
$k_i(Y')\geq k_i(Y)-1\geq 1$.
Finally,
$\epsilon - E(-2a,Y',Y,i',j') = 2a-\epsilon_1\Big(k_j^T(Y')-i'\Big) +
\epsilon_2\Big(k_i(Y)-j'+1\Big)$ vanishes in the limit iff
$k_i(Y)-j'+1=1$, i.e. $j'=k_i(Y)$, what is possible whenever $k_i(Y')\geq k_i(Y)\geq 1$.
Clearly, one gets complementary sets of constraints:
$k_i(Y)\geq k_i(Y')+1$ and $1 \leq k_i(Y)\leq k_i(Y')$, either one or the other
is true, provided $k_i(Y)\geq 1$;
$k_i(Y)\geq k_i(Y')+2$ and $2\leq k_i(Y)\leq k_i(Y')+1$, again either one or
the other is true, this time provided $k_i(Y)\geq 2$.
This means that in the limit of interest the $a$-dependent part of
denominator vanishes as $\epsilon_1$ in the power, equal to twice the number of rows
in $Y$ of the height two or more plus the number of rows in $Y$ of the height exactly
one, i.e.
as $\epsilon_1^{\#(i=1)+\#(i=2)}$.

\paragraph{$\epsilon_1$-dependence of the Nekrasov function: the result.}
Putting all contributions together, one obtains
\be
Z_{Y,Y'}
\sim \frac{\epsilon_1^{2(\#(i=1) + \#(i=2) + \#(i'=1))}}
{\epsilon_1^{\#(i=1)+\#(i'=1)} \epsilon_1^{\#(i=1)+\#(i=2)}}
= \epsilon_1^{\#(i=2) + \#(i'=1)}
\ee
Thus the pair $(Y,Y')$ contributes at $\epsilon_1\rightarrow 0$
iff $\#(i=2) = 1$ and $\#(i'=1)=0$.
This means that $Y'$ should have no rows at all,
i.e. be an empty diagram, while $Y$ should have only height-one
rows,i.e. $Y = [1^{|Y|}]$.
In other words, we proved that
the only Nekrasov functions which do not vanish in the limit are
the {\it chiral} ones with $(Y,Y')=([1^{|Y|}],0)$.

\subsection{Sum of the chiral functions}

It now remains to evaluate these chiral diagrams. This is simple:
\be
Z_{[1^n][0]} = \frac{(-)^n}{n!(\epsilon_1\epsilon_2)^n}
\prod_{i=1}^n \frac{\prod_{f=1}^4 \Big(a+\mu_f + (i-1)\epsilon_1\Big)}
{\Big( 2a+\epsilon +(n-i)\epsilon_1\Big)\Big(2a+(n-i)\epsilon_1\Big)}=
\\
=
\frac{(-)^n}{n!(\epsilon_1\epsilon_2)^n}
\prod_{i=1}^n \frac{\epsilon^{2n}
\epsilon_1^{2n}\Big( \Delta_0+\Delta_1-\Delta_2+(i-1)\Big)
\Big(\Delta_0+\Delta_3-\Delta_4+ (i-1)\Big)+O(\epsilon_1)}
{\Big( 2\alpha_0 +(n-i)\epsilon_1\Big)\Big(2\alpha_0-\epsilon+(n-i)\epsilon_1\Big)}
\longrightarrow\\
\longrightarrow
\frac{1}{n!}
\prod_{i=1}^n \frac{\Big( \Delta_0+\Delta_1-\Delta_2+(i-1)\Big)
\Big(\Delta_0+\Delta_3-\Delta_4+ (i-1)\Big)}
{ 2\Delta_0 +(n-i)}
\ee
The sum
\be\label{final}
\sum_{n=0}^\infty x^n Z_{[1^n][0]} \longrightarrow
\sum_{n=0}^\infty \frac{x^n}{n!}\prod_{k=0}^{n-1}
\frac{(\Delta+\Delta_1-\Delta_2+k)
(\Delta+\Delta_3-\Delta_4+k)}{2\Delta+k}
\ee
which exactly coincides with (\ref{Fhser}).

\subsection{Extra comments}

Some additional comments are in order.
First of all, note that one can repeat the procedure
of this paper in the case of $\epsilon_2\to 0$. Then, since the exchange $\epsilon_1
\leftrightarrow\epsilon_2$ permutes rows and columns in the Young diagrams
in the Nekrasov functions, one represents the {\it same} conformal block (\ref{final})
as a sum of the {\it anti-chiral} partition functions $Z_{[n][0]}$.

Similarly, one can consider alternative possibility of keeping
dimensions  finite in the limit of either $\epsilon_1\to 0$ or $\epsilon_2\to 0$.
Let, say, $\epsilon_1\to 0$. Then instead of (\ref{alim}) one can demand that
$\epsilon -\alpha_i \sim \epsilon_1$, namely that
\be\label{alim2}
\alpha_i = \epsilon -\epsilon_1 \Delta_i + O(\epsilon_1^2)
\ee
All above reasoning works in this case, only the
two Young diagrams in $Z_{[Y],[Y']}$ interchanged, $Y \leftrightarrow Y'$.
If one imposes (\ref{alim2}) in the limit
of $\epsilon_2\to 0$ the  Young diagrams will be
permuted and transposed (the rows and columns exchanged).

In this paper, we considered the limit when the central charge of conformal
theory goes to infinity, while all the dimensions remain finite. Instead, one
could consider the limit when all the dimensions go to infinity along with the
central charge. This is provided just by $\epsilon_1\to 0$ without
adjustment of $\alpha_i$.
To keep conformal blocks finite one should simultaneously take $x \rightarrow 0$.
This limit is much more complicated since
the Nekrasov functions corresponding to several Young diagrams at each level
contribute in this case. Similarly, the analyzes of the conformal side is
more involved in this case: it describes the quasiclassical limit of
conformal theory. It is related to monodromy properties
of an ordinary differential equation
\cite{BPZ,Zam}. A particular case of conformal theory when all $\Delta_i$
remain finite but $\Delta_0$
(which is already quite involved) was considered in \cite{Zam}, the AGT relations
in this case are studied in \cite{mmm3}.

Note that the procedure presented here is immediately continued to higher orders
in $\epsilon_1$: one can construct a perturbation theory and check the AGT
relation order by order in $\epsilon_1$, but in all orders in $x$. This is completely
different from the standard checks of the AGT relation order by order in $x$ but
in all orders in $\epsilon_1$, which were performed in \cite{AGT,Wyl,MMMagt,mmU3}.

The limit $\epsilon_1\rightarrow 0$ corresponds to raising regularization
of the moduli space integral in the {\it instanton} sector, while preserving
it for {\it anti}instantons. In result the integral over the {\it instanton} moduli
space diverges.
Within the "physical" prescription of refs.\cite{Kh,Khrev}, the divergency is compensated
by subtracting a contribution from the
boundary of the moduli space.
This, however, breaks the symmetry of Nekrasov instanton calculus
and, as we now understand, violates the AGT relations, the ones which
would supposedly provide a solid group theory basis for a careful definition
of appropriate $\tau$-functions.
In this letter we suggested to make instanton contribution finite
without subtractions, by taking instead a "double scaling" limit (\ref{alim}).
Of course, in this case instantons continue to dominate over anti-instantons,
so that actually only the instanton contribution survives
in the form of the chiral Nekrasov functions.
The answer, the hypergeometric series implies that the integral
over the {\it instanton} moduli space satisfies a {\it second}-order
differential equation w.r.t. $x$, what is unusual for naive RG equations,
but quite normal for the {\it exact} RG group \cite{exaRG}.
It would be interesting to find a relation of this equation
to the first-order RG equations in Seiberg-Witten theory \cite{RG},
to understand its origin
directly in terms of the instanton calculus, and, probably, even to
extend this relation to the $\epsilon_1\neq 0$ region.

\section{Conclusion}

In this paper we proved the AGT relation for the Virasoro conformal blocks in the
limit of large central charge $c$.
The proof is not conceptual, but simply uses explicit formulas for
both sides of the relation, which can be easily found in this limit.
In fact, it turns out that the limit $c\rightarrow\infty$ imposes
constraints on the conformal blocks and Nekrasov functions which appear
similar to those imposed by
the choice of the {\it special} conditions for external states,
analyzed in
\cite{mmNF,mmU3}: the 4-point conformal block becomes {\it hypergeometric}
and decomposes into a sum of the {\it chiral} Nekrasov functions for diagrams
$([1^n],\emptyset)$.
Anyhow, this fact allows one to extend the existing proof of the AGT relations
from  special to generic external states. Still, it is restricted
to the hypergeometric series, what is now achieved by imposing restriction
on the central charge.

\section*{Acknowledgements}

The work was partly supported by Russian Federal Nuclear Energy
Agency and by RFBR grants 07-02-00878 (A.Mir.),
and 07-02-00645 (A.Mor.).
The work was also partly supported
by joint grants 09-02-90493-Ukr,
09-02-93105-CNRSL, 09-01-92440-CE, 09-02-91005-ANF and by Russian President's Grant
of Support for the Scientific Schools NSh-3035.2008.2.

\end{document}